\documentclass{webofc}
\usepackage[varg]{txfonts}   
\begin{document}
\title{Emergent Higgsless Superconductivity}
%
%

\author{\firstname{M. Cristina} \lastname{Diamantini}\inst{1}\fnsep\thanks{\email{cristina.diamantini@pg.infn.it}} 
\and
        \firstname{Carlo A.} \lastname{Trugenberger}\inst{2}\fnsep\thanks{\email{ca.trugenberger@bluewin.ch}} 
}

\institute{NiPS Laboratory, INFN and Dipartimento di Fisica, University of Perugia, via A. Pascoli, I-06100 Perugia, Italy \and
         SwissScientific, chemin Diodati 10, CH-1223 Cologny, Switzerland 
  }

\abstract{%
We present a new Higgsless model of superconductivity, inspired from anyon superconductivity but P- and T-invariant and generalizable to any dimension. While the original anyon superconductivity mechanism was based on incompressible quantum Hall fluids as average field states, our mechanism involves topological insulators as average field states. In D space dimensions it involves a (D-1)-form fictitious pseudovector gauge field which originates from the condensation of topological defects in compact low-energy effective BF theories.  There is no massive Higgs scalar as there is no local order parameter. When electromagnetism is switched on, the photon acquires mass by the topological BF mechanism. Although the charge of the gapless mode (2) and the topological order (4) are the same as those of the standard Higgs model, the two models of superconductivity are clearly different since the origins of the gap, reflected in the high-energy sectors are totally different. In 2D this type of superconductivity is explicitly realized as global superconductivity in Josephson junction arrays. In 3D this model predicts a possible phase transition from topological insulators to Higgsless superconductors.

}
\maketitle
\section{Introduction}
\label{intro}

The idea of emergence in the interpretation of natural phenomena, as opposed to the reductionist approach, is acquiring more and more importance both in condensed matter and in QCD.
The first example of a physical law that cannot be understood thinking of elementary components but only as a collective emergent phenomenon is the second law of thermodynamics  and irreversibility. 
Quantum chromodynamics (QCD) is the theory of the strong force. However hadron physics is very different from QCD: confinement can be interpreted as an emergent phenomenon at low energy scales. 
But it is in condensed matter that the idea of emergence has found its most striking realisations.
The discovery  of the fractional quantum Hall  (FQHE) \cite{dqhl} effect has revealed the existence of a new state of 
matter characterized by a new type of order: topological order \cite{wen1}. Topological order is a particular type of quantum order
describing zero-temperature properties of a ground state with a gap for all excitations. 
Its hallmark are the degeneracy of the ground state on manifolds
with non-trivial topology, and excitations with fractional spin and statistics, called anyons \cite{wilczek}. The long-distance properties
of these topological fluids are described by Chern-Simons field theories \cite{wen1} with compact gauge group, which break P-and 
T-invariance.  All the striking properties of the FQHE cannot be explained by thinking of elementary components; in fact the  physical properties of this many-body state come from the order or, equivalently, from the organisation of its 
 degrees of freedom in the state.

After Laughlin's discovery of topological quantum fluids, it was conjectured that
a similar mechanism, based on anyon condensation, could be at the origin of  the high-$T_c$ 
superconductivity \cite{wilczek} mechanism for the high-$T_c$ cuprates. The idea, based on the possibility of fractional statistics \cite{wil} in 2 space dimensions (2D), involves fermions interacting with a fictitious, statistical Chern-Simons (CS) gauge field \cite{jac} which turns them into anyons by attaching fictitous magnetic flux to their charge density.
Let us now consider the average field approximation, in which the statistical magnetic field of a uniform particle distribution is substituted by its uniform average $\cal{B}$. As has been extensively discussed in \cite{chen}, this mean field approximation is self-consistent for large integer $k$, although it is widely believed to be valid down to $k=1$ \cite{lau}. 
Suppose that the particle density is $\rho$; the particles will then fall into Landau levels with a filling fraction $\nu = 2\pi \rho /\cal{B}$. But the statistical magnetic field is itself tied to the particle density by the Chern-Simons equation of motion \cite{gerald} implying $\nu = k$. When $k = m \in \mathbb{N}$ exactly $m$ Landau levels are filled and the average field state is gapped. In particular, if we add a real magnetic field $B$ this will disturb the self-consistent Landau level balance and some particles or holes will be excited across the Landau level gap \cite{chen}: this gap for real magnetic fields is essentially the Meissner effect. 
These are not the only gapped states: all states of the Jain hierarchy \cite{jain} of fractional quantum Hall states with filling fraction $\nu = m/(mp+1)$ with $p$ an even integer are also gapped. 

A specific coherent fluctuation of the fermion density {\it together} with the statistical gauge field, however, is gapless. This massless mode is protected from decaying into particle-hole excitations by the average field gap and leads to anyon superfluidity. It can be shown that the origin of the massless mode is not the spontaneous breaking
 of a symmetry but, rather, the necessary restoration of the commutativity of translations which is broken in the average field approximation \cite{gid}. The most studied case is that of two filled Landau levels, corresponding to semions, or half fermions since, in this case, the charge order parameter of the resulting superconductor is 2 \cite{wil}. Unfortunately, the high-$T_c$ cuprates do not exhibit the P and T violation necessarily implied by the superconductivity mechanism based on anyons and thus the idea of anyon superconductivity was quickly abandoned. 

In \cite{higgsless} we revisited the anyon superconductivity mechanism to show that it can be made P- and T-invariant and extended to three space dimensions (3D), where it can be realized upon a phase transition from topological insulators \cite{topins}. To this end we started from the pure gauge formulation of \cite{zee}. We did not consider the usual case of semions (half-fermions) but, rather, we concentrated on the simple case in which the statistical interaction turns the fermions into bosons. 
By soldering together two fermion fluids of opposite spin, interacting with the same fictitious statistical gauge field and filling their respective first Landau levels in the average field approximation, one obtains a single gapless mode with charge 2. The remaining degrees of freedom organize themselves into a massive {\it vector} particle (with two polarisations in 2D) of unit charge, rather than a single neutral, scalar Higgs boson. This massive vector is described by two Chern-Simons terms of opposite parity and the same absolute value of the coupling constant. The P and T symmetries are preserved. 

We showed that this type of P- and T-invariant Higgsless superconductivity can be reformulated by a rotation of the degrees of freedom as a charge and a vortex fluids interacting with each other by a mutual Chern-Simons term. The vortices interact additionally with a pseudovector statistical gauge field which, however, has no self-action. In this formulation, superfluidity arises as follows. In the average field approximation the statistical gauge field provides a uniform charge for the vortices. A charge arises instead of a magnetic field as in standard anyon superconductivity since here the statistical gauge field is a pseudovector. Note, however, that this emergent charge has to be distinguished from the intrinsic charge coupling to real electromagnetic fields, as has been stressed in \cite{moe}. The vortices are subject to the Magnus force \cite{magnus}, which, in 2D, is the exact dual of the Lorentz force and they are thus quantized into (dual) Landau levels. When an integer number of Landau levels is filled the vortices have a gap; this happens in particular for the ground state configuration with no vortices at all. This gap for vortices is nothing else than the Meissner effect. Due to the mutual Chern-Simons term, a generic intrinsic charge fluctuation corresponds to a local variation in the charge distribution felt by the vortices. Exactly as in the original mechanism of anyon superconductivity, this must be accompanied by an excitation of vortex-antivortex pairs through the Landau level gap \cite{chen}. Generic intrinsic charge fluctuation are thus also gapped. If, however the intrinsic charge fluctuation is accompanied by an emergent charge fluctuation such that the total charge felt by vortices remains unchanged, there is no energy price to pay. This is  the superfluid gapless mode. 

When electromagnetism is switched on, the photon acquires mass through the topological Chern-Simons mass mechanism \cite{jac}. The resulting completely gapped state has topological order \cite{wen} characterised by a ground state degeneracy $4^g$ on Riemann surfaces of genus $g$.

The important point is, finally that, in the mutual Chern-Simons formulation, this Higgsless topological superconductivity mechanism is easy to generalize to any number of dimensions \cite{dst2} and, in particular to three space dimensions (3D).  Indeed, a mutual Chern-Simons term can be generalized to any number of dimensions as a topological BF term \cite{birmi}. In 3D, the BF term involves the topological coupling of a standard vector gauge field to a two-form Kalb-Ramond gauge field \cite{kalb} with generalized vector gauge symmetry. Vortices are themselves described by a conserved antisymmetric tensor current and the emergent gauge field coupling to them is also a two-form gauge field. The dual "magnetic field" associated to a pseudotensor gauge field $\alpha_{\mu \nu}$ is the scalar density $F^0$, where $F^{\mu } = \epsilon^{\mu \nu \beta \sigma} \partial_{\nu}\alpha_{\beta \sigma}$. In the average field approximation, this corresponds again to a uniform emergent charge $F^0$ for the vortices. We show that such a uniform Kalb-Ramond scalar density leads to a gap for vortex strings via the Magnus force: this causes the Meissner effect in 3D. The superfluidity mechanism in 3D parallels thus verbatim the 2D situation: vortex strings are gapped in the average field approximation; generic intrinsic charge fluctuations are also gapped since they modify locally the uniform charge distribution felt by the vortices and cause thus the nucleation of gapped vortex strings; a specific combination of intrinsic and emergent charge fluctuations that leaves the scalar Kalb-Ramond density invariant, however, is the gapless superfluid mode. The remaining degrees of freedom represent a massive vector (spin 1) boson via the BF topological mass mechanism \cite{bow}; also in this case there is no scalar Higgs boson. Of course, what does not carry over obviously from 2D, instead, is the original interpretation of the emergent gauge field as a statistics changing interaction. 

The BF term represents the low-energy physics of topological insulators \cite{moore} (both in 2D and 3D). The relation to the original mechanism of anyon superconductivity is thus particularly suggestive. There, the gapped average field state was an incompressible quantum Hall fluid, here it is a topological insulator. Originally, the additional gauge field leading to the gapless superfluid mode was the statistics changing Chern-Simons gauge field, attaching fluxes to charges. Here the additional gauge field is a pseudovector and it attaches emergent charges to intrinsic charges, thereby preserving the P and T symmetries. While such a gauge field can be introduced in any number of dimensions, it cannot arise from fractional statistics in higher dimensions. 

The origin of the emergent gauge field lies, rather in the compactness of the BF term \cite{dst2}. In 2D, the BF term reduces to the mutual Chern-Simons interaction and this represents \cite{dst1} the physics of Josephson junction arrays \cite{faz}. When the two U(1) gauge groups are compact, however, the BF theory has to be formulated as a cutoff theory, e.g. on the lattice, with the unavoidable presence of topological defects. As is well known \cite{kog}, naively irrelevant operators like the Maxwell terms for the two emergent gauge fields can lead to non-perturbative effects if the corresponding masses lie below the cutoff. This is exactly what happens in Josephson junction arrays: the condensation of "electric" topological defects provides the statistical gauge field and global superconductivity in the array is exactly an example of the Higgsless topological superconductivity described above. 

As already pointed out, the BF term represents the low-energy physics of topological insulators \cite{moore}. As an effective low-energy theory of a condensed matter system, it must necessarily be considered as a cutoff theory with compact gauge groups. This implies additional topological excitations in the action. When these are dilute, the resulting physics is indeed that of topological insulators, in which both charges and vortices are screened in the bulk by the topological BF mass $m_{\rm BF}$. When the topological excitations condense, however, there is a phase transition to the Higgsless topological superconductivity described here. In 3D, the Maxwell term for the usual vector gauge field is {\it marginal} and must be added anyhow to the action to establish the complete phase diagram \cite {dst3}. The dynamical term for the Kalb-Ramond gauge field, instead involves an antisymmetric three-tensor. When these terms are included one can easily prove by standard free energy arguments that a phase transition from a topological insulator to a Higgless superconductor takes indeed place when the charge screening in the topological insulator becomes strong enough, $m_{\rm BF}/\Lambda > \lambda_{\rm crit}$, where $\Lambda$ is the ultraviolet cutoff and $\lambda_{\rm crit} \ll 1$. It is not clear to us at this point how the Higgsless superconductors introduced here are related to the "topological superconductors" described e.g. in \cite{topsup}. This is under current investigation. 

We would like to stress that the superconductivity model we presented in \cite{higgsless} is genuinely different than both the BCS dynamical symmetry breaking by pairing and its Abelian Higgs model description. In the present case the components of a "pair" in 2D are already bosons by themselves, having been individually transmuted from fermions by the statistical gauge interaction. It is not the paired fermions that produce a charge condensate via Bose-Einstein condensation, the "condensate" is provided by the emergent charge of the pseudovector statistical gauge field. Neither can our model be phenomenologically described by the Abelian Higgs model, although they share the same charge 2 and the same topological order 4 \cite{han}: in our  model there is no local order parameter and, correspondingly, there is no scalar Higgs boson, but rather a massive charged vector. The properties of our model are, rather, suggestive of a possible connection with Higgsless models of 
 spontaneous symmetry breaking via extra dimensions and the AdS/CFT correspondence \cite{csa}. 

Also note that this model is an example of the fact that one cannot deduce the characteristics of a superconductivity model by looking at the low-energy (photon plus gapless mode) sector exclusively. 
Doing so in this case would have lead to the erroneous conclusion that this model coincides with the abelian Higgs model, since both the charge of the gapless mode and the topological order are the same. Instead, in order to establish the nature of a superconductivity model it is crucial to focus on the mechanism by which the gap is opened and the gapless mode is originated and this is intimately connected with the high-energy sector. In a sense, all possible superconductivity models must have the same phenomenological low-energy structure, it is only the origin of the gap, and thus the high-energy physics, that can distinguish them.

\section{Mutual anyon superconductivity}
\label{sec-1}

We will not present here the detailed analysis of the model described above, that is given in \cite{higgsless}. We will instead present a particularly simple and familiar form of this  Higgsless superconductivity mechanism that can be easily generalised to any number of dimensions, given by the Lagrangian 

\begin{equation} 
{\cal L} = {1\over 4\pi} {\cal M}^T_{\mu} \Lambda_M \ \epsilon^{\mu \nu \sigma} \partial_{\nu}{\cal M}_{\sigma} + 2 A_{\mu } j^{\mu} \ ,
\label{ctwo}
\end{equation}
with 
\begin{equation}
{\cal{M}}_{\mu} = (a_{\mu}, b_{\mu },  -\alpha_{\mu}) \ ,
\end{equation}
and
\begin{equation}
\Lambda_M= \begin{pmatrix}
  0 & p+1 &  1  \\
  p+1 & 0 & 0 \\
  1 & 0 & 0  \\
\end{pmatrix} \ .
\label{cthree}
\end{equation}
It involves only mutual Chern-Simons terms,
\begin{equation}
{\cal{L}} = {p+1\over 2\pi} a_{\mu} \epsilon^{\mu \nu \sigma} \partial_{\nu} b_{\sigma} + {1\over 2\pi} a_{\mu} \epsilon^{\mu \nu \sigma} \partial_{\nu} \alpha_{\sigma} + 2 A_{\mu} j^{\mu} \ .
\label{ctwo}
\end{equation} 
and describes specifically a mutual Chern-Simons interaction between vortices with conserved current 
\begin{equation}
\phi^{\mu} = {1\over 2\pi} \epsilon^{\mu \nu \sigma} \partial_{\nu} a_{\sigma} \ ,
\label{cthree}
\end{equation}
and charges with conserved current 
\begin{equation}
j^{\mu} = {1\over 2\pi} \epsilon^{\mu \nu \sigma} \partial_{\nu} b_{\sigma} \ ,
\label{cfour}
\end{equation}
and charge unit 2. In addition there is also a mutual Chern-Simons interaction between the vortices and the statistical gauge field. Since $a_{\mu}$ and $b_{\mu}$ are vector and pseudovector gauge fields, respectively and the statistical gauge field $\alpha_{\mu}$ is also a pseudovector, this Lagrangian respects both the discrete symmetries P and T. 

In this formulation, the superconductivity mechanism is very simple to illustrate. In the average field approximation, the statistical gauge fields provides a uniform emergent charge for the vortices. Keep in mind that this emergent charge is different from the intrinsic charge that couples to $A_{\mu}$. Due to the Magnus force, which is the exact dual of the Lorentz force in 2D, 
vortices fall into Landau levels and any configuration with a completely filled Landau level has a gap. This is valid, in particular, for the vacuum with all Landau levels empty, i.e. the state with no vortices. This gap for vortices represents the Meissner effect. Generic intrinsic charge fluctuations are also gapped since, due to the mutual Chern-Simons term, they distort the uniform emergent background charge field for the vortices, and must thus be accompanied by vortex or anti-vortex nucleation, exactly the same mechanism as in traditional anyon superconductivity \cite{chen}. There is however a particular combination of 1 intrinsic charge (with charge unit 2) and $(p+1)$ emergent charges that can fluctuate freely without altering the overall charge distribution felt by the vortices: this is the isolated gapless mode leading to superfluidity. This implies that low-energy emergent charges carry intrinsic charge $2/(p+1)$. The relative Aharonov-Bohm statistical phase acquired by one such emergent charge when it is carried around a vortex is thus $\pi (p+1)$. This is compatible with the relative statistical phase $\pi $ of one intrinsic charge 2 only if $p$ is an even integer. In other words, only if $p$ is an even integer, the emergent charge can be attached to intrinsic charges with charge unit 2 without altering the real flux quantization in units of $\pi$. 

In these variables the relation to the original mechanism of anyon superconductivity becomes fully exposed. If we freeze fluctuations of the emergent gauge field $\alpha$ in (\ref{ctwo}) we are left with the average field approximation in which both vortices and intrinsic charges are gapped and with the effective action given by the first mutual Chern-Simons term in (\ref{ctwo}). For $p=0$ this is nothing else than the effective action for a 2D topological insulator \cite{moore}, which describes exactly a state in which both vortices and charges are fully screened. While the gapped average field state of anyon superconductivity was an incompressible quantum Hall state, in the present model it is a topological insulator. The additional gauge field that leads to an isolated superfluid mode in anyon superconductivity was the statistics changing Chern-Simons field, carrying vorticity and thus breaking the P and T symmetries. In the present mode, the mechanism leading to an isolated superfluid mode is exactly the same: the additional gauge field however is a pseudovector that has its origin in the condensation of topological defects, as we show below. It carries a scalar emergent charge and thus does not break P ans T. 

Having discussed in detail how the gapless superfluid mode arises, let us now turn to the effects of the electromagnetic coupling in (\ref{ctwo}). We have already seen in the previous section that this causes the entire model to become gapful via the BF topological mass mechanism \cite{bow}. This, on the other side, goes hand in hand with topological order \cite{wen}, characterised by a ground state degeneracy $d^g$ on  Riemann surfaces of genus $g$, when the gauge fields are compact, as in the present case.  The degeneracy parameter $d$ for multi-component Chern-Simons terms $(1/4\pi) {\bf a}_{\mu} M \epsilon^{\mu \nu \sigma} \partial_{\nu} {\bf a}_{\sigma}$  is governed by the determinant of $M$ \cite{poly}. If $M$ has only integer entries, then $d=|{\det } \ M|$. If the entries are rational one has to construct a representation $M=M_1 M_2^{-1}$ where $M_1$ and $M_2$ have mutually prime integer entries. The degeneracy parameter is then $d=|{\det } \ M_1 \ {\rm det}\ M_2| = 4$ in the present case \cite{higgsless}.

To complete the model we should also give a picture of how superconductivity is destroyed at temperatures approaching the high-energy vector mass. To this end we recall that the mass of the vector particle is of the order of the gap for vortices in the average field approximation. At temperatures comparable with this energy scale vortices can cross this gap and become thus liberated. We expect thus the transition to be due to the deconfinement of vortices, which, in 2D is a Kosterlitz-Thouless transition \cite{kos} and, as expected is not related to the restoration of a symmetry, nor does it involve an order parameter. This is fully confirmed experimentally. Indeed, in 2D, our model of topological Higgsless superconductivity is explicitly realized \cite{dst2} as global superconductivity in Josephson junction arrays \cite{faz}. In these granular materials, superconductivity is known to be destroyed at high temperatures due to the deconfinement of vortices in a Kosterlitz-Thouless phase transition \cite{faz}: no spontaneous symmetry breaking is involved. 

Let us recapitulate the properties of this novel superconductivity mechanism. At the superfluid level there is one isolated gapless mode of charge 2 but no Higgs boson; rather, there is a charged, massive Chern-Simons vector. When electromagnetism is switched on, the photon acquires mass by combining with the gapless mode via the topological BF mass mechanism \cite{jac, bow}. The resulting topological order is 4. Although the charge 2 of the gapless mode and the topological order $d=4$ coincide with those of the abelian Higgs model \cite{han}, this is quite different than the standard BCS pairing mechanism. Here, the individual components of a charge 2 "pair" are already bosons, having been statistically transmuted from fermions by an emergent statistical gauge field. Nor can this model be phenomenologically described by the Abelian Higgs model: there is no local order parameter and, correspondingly, there is no Higgs boson. The properties of the present model are, rather suggestive of a possible connection with Higgsless models of spontaneous symmetry breaking via extra dimensions and the AdS/CFT correspondence \cite{csa}. 

Also note that this model is a paradigm example of the fact that one cannot deduce the characteristics of a superconductivity model by looking at the low-energy (photon plus gapless mode) sector exclusively. 
Doing so in this case would have lead to the erroneous conclusion that this model coincides with the abelian Higgs model, since both the charge of the gapless mode and the topological order are the same. Instead, in order to establish the nature of a superconductivity model it is crucial to focus on the mechanism by which the gap is opened and the gapless mode is originated and this is intimately connected with the high-energy sector. In a sense, all possible superconductivity models must have the same phenomenological low-energy structure, it is only the origin of the gap, and thus the high-energy physics, that can distinguish them. In the present case, the gapless mode is intimately tied to the emergent statistical gauge field. We must thus clarify the origin of this gauge field. This is the subject of the next section.

\section{The emergent gauge field as a topological excitation}
\label{sec-2}

The mutual Chern-Simons interaction between charges and vortices, first term in equation (\ref{ctwo}), is easily generalisable \cite{dst2} to any number of dimensions as a BF term \cite{birmi},
\begin{equation}
S_{BF} = {k \over 2 \pi} \int_{M_{d+1}} a_1 \wedge d b_{d-1} \ ,
\label{done}
\end{equation}
on a manifold of spatial dimension $d$. Here $a_1$ is a one-form and, correspondingly, $b_{d-1}$ is a $(d-1)$-form. The conserved current $j_1 = * db_{d-1}$ represents charge fluctuations, while the generalized current $\phi_{d-1} = * da_1$ describes conserved fluctuations of $(d-2)$-dimensional branes. The form $b_{d-1}$ is a pseudo-tensor, while $a_1$ is a vector: the BF coupling is thus P- and T-invariant. The BF term always represents a mass term for the gauge fields $a_1$ and $b_{d-1}$ \cite{dst2}. In the special case $d=2$, it reduces exactly to the mutual Chern-Simons term in (\ref{ctwo}). 

The action (\ref{done}) has the usual gauge symmetry under shifts 
\begin{equation}
a_1\rightarrow a_1+ \xi_1 \ ,
\label{dtwo}
\end{equation}
with $\xi_1$ a closed 1-form, $d \xi_1 = 0$, provided vanishing boundary conditions for the corresponding field strength are chosen. However, it has also a generalized Abelian gauge symmetry under transformations
\begin{equation}
b_{d-1} \rightarrow b_{d-1} + \eta_{d-1} \ ,
\label{dthree}
\end{equation}
where $\eta_{d-1}$ is a closed $(d-1)$-form: $d \eta_{d-1} = 0$. The important point is that, in application to low-energy effective models of condensed matter systems, these gauge symmetries have to be considered as {\it compact}. As is well known \cite{polya}, the compactness of the gauge fields leads to the presence of topological defects. In the present case there are both magnetic topological defects, associated with the compactness of the usual gauge symmetry (\ref{dtwo}) and electric ones, associated with the compactness of the gauge symmetry (\ref{dthree}). The electric topological defects  couple to the form $a_1$ and are string-like objects  described by a singular, closed 1-form $Q_1$: they describe the world lines of point charges. Magnetic topological 
defects couple to the form $b_{d-1}$ and are closed $(d-1)$-branes described  by a singular $(d-1)$-dimensional form $\Omega_{d-1}$. In 2 space dimensions they also reduce to string-like objects that describe the world lines of point vortices. 
These forms represent the singular parts of the field strenghts $da_1$ and $db_{d-1}$, allowed by the compactness of the gauge
symmetries \cite{polya}, and are such that the integral of their Hodge dual  over any  hypersurface of dimensions d and 2, respectively, is $2 \pi n$ with $n$ an integer.  In an effective field theory approach they have structure on the
scale of the ultraviolet cutoff. 

Thus far we have described the kinematics of topological defects. The dynamics depends, of course on all terms present in the full action. Concretely, however, topological defects can be dilute, in which case they do not have any effect, or can condense. The phase with condensed topological defects has a completely different character. In this section we shall not discuss the conditions for condensation of topological defects but we will instead focus on the characteristics of the phase in which the electric topological defects condense. 

A formal derivation of the action with condensed topological defects requires the introduction of an ultraviolet regularization, e. g. in the form of a lattice gauge theory. The result of this procedure \cite{polya}, however, amounts to promote the form $Q_1$ to a dynamical field over which one has to sum in the partition function, 
\begin{eqnarray}
&Z = \int {\cal D}a_1 {\cal D}b_{d-1} {\cal D}Q_1 \nonumber \\ 
&\exp \left[ i {k \over 2 \pi}\int_{M_{d+1}} \left( a_1 \wedge d b_{d-1} + a_1 \wedge *Q_1 \right) \right] \ .
\label{dfour}
\end{eqnarray}
Since $Q_1$ is closed, one can represent it as $Q_1 = d \alpha_{d-1}$. The summation over $Q_1$ in the partition function can then be substituted with a summation over $\alpha_{d-1}$ provided the gauge volume due to the additional symmetry $\alpha_{d-1} \to \alpha_{d-1} + \lambda_{d-1}$ with $\lambda_{d-1}$ a closed $(d-1)$ form, is duly subtracted. The resulting model in the electric condensation phase can thus be formulated in terms of the three dynamical gauge fields $a_1$, $b_{d-1}$ and $\alpha_{d-1}$ and has the action
\begin{equation}
S = {k \over 2 \pi}\int_{M_{d+1}} \left( a_1 \wedge d b_{d-1} + a_1 \wedge d\alpha_{d-1} \right)  \ .
\label{dfive}
\end{equation}
This is the generalization to any dimensions of the model (\ref{ctwo}) for the case $p=0$. In 2 space dimensions, $d=2$, it reduces exactly to the $p=0$ version of (\ref{ctwo}). The scalar gapless mode is represented by the gauge field combination $\varphi_{d-1}= b_{d-1}-\alpha_{d-1}$, with corresponding conserved current $j_1 = *d\varphi_{d-1}$. The remaining $d$ massive degrees of freedom in $a_1$ and $b_{d-1}+\alpha_{d-1}$ represent the massive vector in $d$ space dimensions. 

This shows that the origin of the emergent gauge field in this superconductivity model lies in the condensation of topological defects in effective, compact BF field theories of topological matter. It is no wonder that there is no Higgs boson in this model: there is no local order parameter, superconductivity arises as a consequence of a condensation of topological defects and the two phases can be distinguished only by the behaviour of Wilson loops \cite{polya,sim}. Notice that, due to the BF mass term, the topological defects $Q_1$ have short-range interactions: in a Euclidean formulation, their self-energy is proportional to the length of their world lines, exactly as their entropy. The condensation (or lack thereof) is thus determined by an energy-entropy balance. Since it arises due to the condensation of topological defects without any local order parameter, we find it appropriate to call this superconductivity model "topological". 

In two space dimensions this topological Higgsless superconductivity is explicitly realized \cite{dst1} as global superconductivity in Josephson junction arrays \cite{faz}. The 3D case is perhaps even more interesting, as we now show. 

\section{3D: turning a topological insulator into a superconductor} 
\label{sec-3}
In three space dimensions the BF term (\ref{done}) is again the low-energy effective action for topological insulators \cite{moore}. These are topological states of matter in which both charges and vortices are completely screened in the bulk, but which support metallic edge states \cite{topins}. Our results imply that, if topological defects condense, topological insulators could turn into Higgsless topological superconductors described by the Lagrangian
\begin{equation}
{\cal{L}} = {1\over 2\pi} a_{\mu} \epsilon^{\mu \nu \sigma \rho} \partial_{\nu} b_{\sigma \rho} + {1\over 2\pi} a_{\mu} \epsilon^{\mu \nu \sigma \rho} \partial_{\nu} \alpha_{\sigma \rho} \ ,
\label{eone}
\end{equation} 
where we have left out, for simplicity, the electromagnetic coupling. Here $b_{\mu \nu}$ and the emergent gauge field $\alpha_{\mu \nu}$ are Kalb-Ramond antisymmetric (two-form) gauge fields \cite{kalb}, with generalized gauge invariance under the transformations
\begin{eqnarray}
b_{\mu \nu} &&\to b_{\mu \nu} + \partial_{\mu} \lambda_{\nu}-\partial_{\nu} \lambda_{\mu} \ ,
\nonumber \\
\alpha_{\mu \nu} &&\to \alpha_{\mu \nu} + \partial_{\mu} \eta_{\nu}-\partial_{\nu} \eta_{\mu} \ .
\label{etwo}
\end{eqnarray}
The dual field strength 
\begin{equation}
j^{\mu}= {1\over 2\pi} \epsilon^{\mu \nu \sigma \rho} \partial_{\nu}b_{\sigma \rho} \ ,
\label{ethree}
\end{equation}
of the Kalb-Ramond field is a vector field (since the emergent gauge field is a pseudotensor) which represents the charge current, while the current of vortex strings is given by the usual antisymmetric dual field strength
\begin{equation}
\phi^{\mu \nu} = {1\over 2\pi } \epsilon^{\mu \nu \sigma \rho} \partial_{\sigma} a_{\rho} \ . 
\label{efour}
\end{equation}

The superconductivity mechanism in three dimensions parallels exactly the 2D mutual anyon superconductivity presented above, as we now show.  In the average field approximation, the emergent gauge field provides a uniform emergent charge given by the Kalb-Ramond dual field strength $F^0 = \epsilon^{ijk} \partial_i \alpha_{jk}$ . This uniform Kalb-Ramond emergent charge causes a gap for vortices via the Magnus force. To see this, let us consider an elementary vortex string with world-surface parametrized by ${\bf x}(\xi_0, \xi_1)$ and action
\begin{equation}
S_v=\int d^2 \xi  \ T\ \sqrt{g} g^{ab} {\cal D}_a x_{\mu} {\cal D}_b x^{\mu} + \alpha_{\mu \nu}  \epsilon^{ab} \partial_a x^{\mu} \partial_b x^{\nu}  \ ,
\label{esix}
\end{equation}
where ${\cal{D}}_a$ are the covariant derivatives with respect to the induced metric $g_{ab}=\partial_a x_{\mu} \partial_b x^{\mu}$, $g$ is the determinant of this induced metric and $2T$ is the (bare) string tension. The first term in this action is the celebrated Polyakov action \cite{polya} whereas the second term represents the Magnus force coupling to the antisymmetric Kalb-Ramond emergent gauge field. 

We analyze this model along the lines of \cite{klei} by introducing a Lagrange multiplier $l^{ab}$ to enforce the constraint $g_{ab}= \partial_a x_{\mu} \partial_b x^{\mu}$ and extending the action (\ref{esix}) to 
\begin{equation}
S_v \to S_v + \int d^2 \xi \sqrt{g} \ l^{ab} \left( \partial_a x_{\mu} \partial_b x^{\mu} - g_{ab} \right) \ . 
\label{eseven}
\end{equation}
We then parametrize the world-surface in a Gauss map by choosing to set the coordinate "1" of space-time along the vortex, $x_{\mu} (\xi_0, \xi_1) = \left( \xi_0, \xi_1, \phi^i (\xi_0, \xi_1) \right)$, where $\phi^i (\xi_0, \xi_1)$, $i=2,3$, describe the 2 transverse fluctuations. With the usual homogeneity and isotropy ansatz $g_{ab} = \rho \ \eta_{ab}$, $l^{ab} = l \ g^{ab}$ 
we obtain
\begin{equation}
S=\int d^2 \xi \ \theta +  \int d^2 \xi  \ {T_r\over 2} \partial_a \phi^i \partial^a\phi^i + \alpha_{\mu \nu}  \epsilon^{ab} \partial_a x^{\mu} \partial_b x^{\nu}
\label{eeight}
\end{equation}
where $\theta =T_r -2l\rho$ and $T_r = 2(T+l)$ is the renormalized string tension. At this point we can partially fix the gauge for the antisymmetric Kalb Ramond gauge field by choosing the partial Weyl gauge conditions $\alpha_{02} = 0$, $\alpha_{03} = 0$ and the partial axial gauge condition $\alpha_{23}=0$. This gives 
\begin{equation}
S=\int d^2 \xi \ {T_r\over 2} \ \partial_a \phi^i \partial^a\phi^i - \theta_0 - \theta_i \dot \phi^i \ ,
\label{enine}
\end{equation}
where we have omitted the first constant term (irrelevant for the following) and 
\begin{equation}
\theta _0 = \alpha_{10}\ , \qquad \theta_i = \alpha_{1i} \ , \qquad i=2,3 \ , 
\label{eten}
\end{equation}
are the components of an effective $(2+1)$-dimensional gauge field with residual gauge invariance under transformations 
\begin{eqnarray}
\theta_0 &&\to \theta_0 - \partial_0 \lambda_1 \ ,
\nonumber \\
\theta_i && \to \theta_i - \partial_i \lambda_1 \ ,
\label{eeleven}
\end{eqnarray}
where $\lambda_1$ is the first component of the original vector gauge parameter of the two-form Kalb-Ramond gauge field, which embodies the only residual gauge freedom in this gauge.  The Hamiltonian corresponding to (\ref{enine}) is 
\begin{equation}
H = \int d\xi_1 \ {1\over 2T_r} \left(  P^i-  \theta^i \right)^2 + {T_r\over 2} \partial_{\xi_1}\phi^i \partial_{\xi_1}\phi^i  + \theta_0 \ ,
\label{etwelve}
\end{equation}
where $P^i = T_r \dot \phi^i + \theta^i$ is the canonical momentum density. This Hamiltonian is equivalent to a continuous sequence labeled by $\xi_1$ of particles, held together by the elastic term ${T_r\over 2} \partial_{\xi_1}\phi^i \partial_{\xi_1}\phi^i$ and subject to an effective $(2+1)$-dimensional electromagnetic potential $\theta_{\mu}$. In the gauge we have chosen, a uniform Kalb-Ramond dual field strength $F^0 = \epsilon^{ijk} \partial_i \alpha_{jk}$ is equivalent to a uniform, effective $(2+1)$-dimensional emergent charge (dual magnetic field) $F^0 = \partial_3 \theta_2 - \partial_2 \theta_3$. Each "particle" in the sequence constituting the vortex feels thus a uniform emergent charge due to the Magnus force. The ground state energy of a vortex is thus given by the sum of the zero-point energies of a sequence of Landau oscillators, $E = (F^0/2T_r) (L/\zeta)$, where $L$ is the length of the vortex and $\zeta$ the ultraviolet cutoff. As a consequence, vortices are gapped in presence of a uniform Kalb-Ramond dual field strength (emergent charge): this gap leads to the Meissner effect. 

Exactly as in the 2D case, generic intrinsic charge fluctuations are also gapped, since a generic fluctuation in the current (\ref{ethree}) causes a distortion in the total charge distribution felt by the vortices, with the consequent nucleation of vortices or anti-vortices, which costs energy, as we have just shown above. There is however one particular combination $\varphi_{\mu \nu} = b_{\mu \nu} - \alpha_{\mu \nu}$ of intrinsic and emergent charges that can oscillate coherently without modifying the total background charge distribution felt by the vortices: this is the isolated gapless mode implying superfluidity in the model (\ref{eone}) and superconductivity when it is coupled to electromagnetic fields. The remaining 3 degrees of freedom $a_{\mu}$ and $b_{\mu \nu} + \alpha_{\mu \nu}$ are coupled by a topological BF term as is evident from (\ref{eone}). As is well known \cite{bow} they represent thus a massive vector (spin 1) particle. Also in this case there is no scalar Higgs boson and no local order parameter. 

As we have already pointed out, the first term in (\ref{eone}) is the low-energy effective field theory for topological insulators \cite{moore}. As in 2D, thus, the average field state for our superconductivity mechanism is a fully gapped topological insulator. Actually, our results imply that a 3D topological insulator is the 3D analogue of a full first Landau level when a Kalb-Ramond scalar density $F^0$ is applied to vortex strings instead than a 2D pseudoscalar magnetic field to charged particles. 
Our results imply, thus, that the condensation of the emergent gauge field $\alpha_{\mu \nu}$ could thus turn a topological insulator into a topological Higgsless superconductor. But what could drive this phase transition?  The original idea of formulating low-energy effective field theories of condensed matter systems in terms of emergent gauge fields was based on the fact that, typically, the dynamical terms for the gauge fields are infrared irrelevant and, thus, the whole physics is determined by the topological terms. First of all, in 3D, the usual Maxwell term for the gauge field $a_{\mu}$ is marginal and must anyhow be included in the action. Secondly, as is well known \cite{kog}, when the theory is considered as a cutoff theory, as it must in the present context,  this argument is valid only in the perturbative scaling regime in which all masses of irrelevant operators are beyond the ultraviolet cutoff. In the opposite regime, however, when there are masses of high-energy fields in the observable regime below the cutoff, these can induce non-perturbative effects, like phase transitions. This is exactly what happens in this model, as we now show. 

Let us do so in the (relativistic) Euclidean space formulation of the model with the dynamical gauge field terms added to the topological BF term
\begin{eqnarray}
S_E &&= \int d^3 x {1\over 4e^2} f_{\mu \nu} f_{\mu \nu} -  {i\over 2\pi} a_{\mu} \epsilon^{\mu \nu \sigma \rho} \partial_{\nu} b_{\sigma \rho} +
\nonumber \\
&&+ {1\over 12g^2} h_{\mu \nu \alpha } h_{\mu \nu \alpha} - {i\over 2\pi} a_{\mu} Q_{\mu}\ , 
\label{ethirteen}
\end{eqnarray}
where we have included the topological defects $Q_{\mu}$ due to the compactness of the $BF$ terms, which in Euclidean space become strings corresponding to the Minkowski world lines of charges and 
\begin{eqnarray}
f_{\mu \nu} &&= \partial_{\mu}a_{\nu}-\partial_{\nu}a_{\mu} \ ,
\nonumber \\
h_{\mu \nu \sigma} &&= \partial_{\mu } b_{\nu \sigma} + \partial_{\nu } b_{\sigma \mu} + \partial_{\sigma } b_{\mu \nu} \ .
\label{efourteen}
\end{eqnarray}
The first term in (\ref{ethirteen}) is the usual Maxwell term for the gauge field $a_{\mu}$, the third one is the dynamical term for the three-form Kalb-Ramond field strength $h_{\mu \nu \sigma}$. Recalling that the dual Kalb-Ramond field strength $f_{\mu} = (1/6) \epsilon_{\mu \nu \sigma \rho} \partial_{\nu}b_{\sigma \rho} = \pi j_{\mu}/3$ coincides with the conserved charge current, eq. (\ref{ethree}), we see that the Kalb-Ramond dynamical term represents a charge-charge interaction with strength $g$ of dimension mass. Correspondingly, since $\tilde f_{\mu \nu} = 4\pi \phi_{\mu \nu}$ represents the conserved vortex current, eq. (\ref{efour}), the Maxwell term embodies a vortex-vortex interaction with dimensionless strength $e$. 

We can now integrate out the gauge fields $a_{\mu}$ and $b_{\mu \nu}$ to obtain en effective action for the topological excitations $Q_{\mu}$. As we have explained in the previous section, however, a proper treatment of the topological excitations requires the introduction of an ultraviolet cutoff, by formulating the model, e.g. on a lattice. A fully self-contained derivation of the lattice effective action for the topological excitations is beyond the scope of the present paper. We have presented the complete calculation elsewhere \cite{dst1, dst3}, here we simply quote the result, which is the evident lattice translation of the result one would obtain from the continuous action (\ref{ethirteen}). The partition function is given by
\begin{eqnarray}
Z_{\rm top} &&= \sum_{\left\{ Q_{\mu} \right\}} {\rm exp} \left( -S_{\rm top} \right) \ ,
\nonumber \\
S_{\rm top} &&= \sum_{{\bf x}} {e^2\over 2 l^2} \ Q_{\mu} {\delta_{\mu \nu} \over m^2-\nabla^2} Q_{\nu} \ ,
\label{efifteen}
\end{eqnarray}
where $l$ is the lattice spacing  and $m=eg/\pi$ is the topological BF mass. This mass causes the screening of both charges and vortices in the topological insulator phase. As expected, the topological excitations have also short-range interactions on the scale of this mass. The phase structure implied by this result can be derived by the usual free energy arguments \cite{kle2}.
By approximating the short-range interactions with contact terms one can assign an energy $ (e^2/2l^2m^2) L$ to a topological excitation of length $L$. The entropy of a string of length $L$ is also proportional to $L$, $\mu L$, where $\mu \simeq {\rm ln} 7$ since, in 4 Euclidean dimensions a string has 7 directions to choose from, without backtracking. The free energy of a string of length $L$ is thus given approximately by
\begin{equation}
F = \left( {e^2\over 2 (lm)^2} - \mu \right) \ L \ .
\label{esixteen}
\end{equation}
When $ml < e/\sqrt{2\mu}$ the free energy is positive, hence it is minimised by strings of length 0 or, in other words, topological excitations are dilute: this is the topological insulator phase. If instead $ml > e/\sqrt{2\mu}$ the free energy is dominated by the entropy and becomes negative, hence long strings are favoured and topological excitations condense: this is the superconductor phase. This shows that, in a fully non-perturbative treatment including topological excitations, topological insulators develop a transition to Higgsless topological superconductivity when the range of the screened Coulomb interaction becomes smaller than a critical value $(\sqrt{2\mu}/e) l$. 

\section{Conclusions}
\label{sec-4}
The anyon superconductivity mechanism, in its P- and T-invariant doubled formulation, can be extended to any number of dimensions. In 3D it involves a compact topological BF term between a usual vector gauge field and a two-form Kalb-Ramond gauge field. The emergent gauge field is also a two-form gauge field arising from the condensation of topological defects. The basic mechanism is the same as in 2D: in the average field approximation a uniform Kalb-Ramond emergent charge causes a gap for vortex strings via the Magnus force. The gapped average field state is a topological insulator. One particular combination of intrinsic and emergent charges, however, is gapless and leads to superfluidity. There is no local order parameter and, thus no Higgs scalar, rather the massive mode is a charged vector (spin 1) particle. This mechanism predicts a possible phase transition from topological insulators to Higgsless superconductors if the charge screening in the topological insulator b
 becomes strong enough. 

M.C. Diamantini  acknowledges financial support from
the European Commission (FPVII, Grant agreement no:
318287, LANDAUER and Grant agreement no: 611004, ICT-Energy) and ONRG grant N00014-11-1-0695.

%
%

\end{document}